\documentclass[aps,prb,reprint,superscriptaddress,showpacs,longbibliography,amsmath,amssymb]{revtex4-1}

\usepackage{graphicx}
\usepackage{color}
\usepackage{miller}
\usepackage{hyperref}
\usepackage{graphicx}
\usepackage{color}

\usepackage{siunitx}
\DeclareSIUnit\Torr{Torr}
\DeclareSIUnit\BL{BL}

\begin{document}

\title{Bi monocrystal formation on InAs(111)A and B substrates}

\author{L.~Nicola\"{\i}}
\affiliation{New Technologies-Research Center, University of West Bohemia, Univerzitni~8, 306\,14 Pilsen, Czech Republic}

\author{J.-M.~Mariot}
\affiliation{Sorbonne Universit\'e, CNRS (UMR~7614),\\ 
Laboratoire de Chimie Physique--Mati\`ere et Rayonnement, 4 place Jussieu, 75252~Paris~Cedex~05, France}
\affiliation{Synchrotron SOLEIL, L'Orme des Merisiers, Saint-Aubin, BP~48, 91192 Gif-sur-Yvette, France}

\author{U.~Djukic}
\affiliation{Laboratoire de Physique des Mat\'eriaux et des Surfaces, Universit\'e de Cergy-Pontoise, 5 mail Gay-Lussac, 95031~Cergy-Pontoise, France}

\author{W.~Wang}
\affiliation{Link\"{o}ping University, Department of Physics, Chemistry and~Biology~/ Surface and Semiconductor Physics, 581\,83 Link\"{o}ping, Sweden}

\author{O.~Heckmann}
\affiliation{Laboratoire de Physique des Mat\'eriaux et des Surfaces, Universit\'e de Cergy-Pontoise, 5 mail Gay-Lussac, 95031~Cergy-Pontoise, France}
\affiliation{DRF, IRAMIS, SPEC -- CNRS/UMR~3680, B\^{a}t.~772, L'Orme des Merisiers, CEA Saclay, 91191 Gif-sur-Yvette Cedex, France}

\author{M.~C.~Richter}
\affiliation{Laboratoire de Physique des Mat\'eriaux et des Surfaces, Universit\'e de Cergy-Pontoise, 5 mail Gay-Lussac, 95031~Cergy-Pontoise, France}
\affiliation{DRF, IRAMIS, SPEC -- CNRS/UMR~3680, B\^{a}t.~772, L'Orme des Merisiers, CEA Saclay, 91191 Gif-sur-Yvette Cedex, France}

\author{J.~Kanski}
\affiliation{Chalmers University of Technology, Department of Physics, 412\,96~Gothenburg, Sweden}

\author{M.~Leandersson}
\affiliation{MAX~IV Laboratory, Lund University, P.O.~Box~118, 221\,00~Lund, Sweden}

\author{J.~Sadowski}
\affiliation{Lund University, MAX-lab, P.O.~Box~118, 221\,00~Lund, Sweden}
\affiliation{Institute of Physics, Polish Academy of Sciences, al.\ Lotnik\'ow 32/46, 02-668 Warszawa, Poland}

\author{T.~Balasubramanian}
\affiliation{MAX~IV Laboratory, Lund University, P.O.~Box~118, 221\,00~Lund, Sweden}

\author{I.~Vobornik}
\affiliation{Istituto Officina dei Materiali, TASC~Laboratory, CNR, Area Science Park - Basovizza, Strada Statale~14, km~163.5, 34149~Trieste, Italy}

\author{J.~Fujii}
\affiliation{Istituto Officina dei Materiali, TASC~Laboratory, CNR, 34014~Trieste, Italy}

\author{J. Braun}
\affiliation{Ludwig-Maximilians-Universit\"{a}t M\"{u}nchen, Department Chemie, Butenandtstra{\ss}e~11, 81377~M\"{u}nchen, Germany}
 
\author{H. Ebert}
\affiliation{Ludwig-Maximilians-Universit\"{a}t M\"{u}nchen, Department Chemie, Butenandtstra{\ss}e~11, 81377~M\"{u}nchen, Germany}

\author{J.~Min\'{a}r}
\affiliation{New Technologies-Research Center, University of West Bohemia, Univerzitni~8, 306\,14 Pilsen, Czech Republic}

\author{K.~Hricovini}
\affiliation{Laboratoire de Physique des Mat\'eriaux et des Surfaces, Universit\'e de Cergy-Pontoise, 5 mail Gay-Lussac, 95031~Cergy-Pontoise, France}
\affiliation{DRF, IRAMIS, SPEC -- CNRS/UMR~3680, B\^{a}t.~772, L'Orme des Merisiers, CEA Saclay, 91191 Gif-sur-Yvette Cedex, France}

\date{\today}

\begin{abstract}
The growth of Bi films deposited on both A and B~faces of InAs\hkl(111) has been investigated by low-energy electron diffraction, scanning tunneling microscopy, and photoelectron spectroscopy using synchrotron radiation. The changes upon Bi deposition of the In~$4d$ and Bi $5d_{5/2}$ photoelectron signals allow to get a comprehensive picture of the Bi/InAs\hkl(111) interface. From the initial stages the Bi growth on the A face (In-terminated InAs) is epitaxial, contrary to that on the B face (As-terminated InAs) that proceeds via the formation of islands. 
Angle-resolved photoelectron spectra show that the electronic structure of a $\approx \SI{10}{\BL}$ deposit on the A face is identical to that of bulk~Bi, while more than $\approx 30$~BL are needed for the B face. Both bulk and surface states are well accounted for by fully relativistic $ab$ $initio$ spin-resolved
photoemission calculations.
\end{abstract}


\maketitle

\section{Introduction}
. 

A number of recently discovered materials, such as topological insulators (TIs) or Weyl semimentals, share a fundamental similarity. Rather than fermions obeying the usual Schr\"{o}dinger Hamiltonian, their low-energy fermionic excitations behave as massless Dirac particles. This emergent property of fermions in condensed matter systems defines the unifying framework for a class of materials we call ``Dirac material''. The emergence of Dirac excitations is controlled by symmetries of the material, such as time-reversal, gauge, and spin-orbit coupling.

The family of TIs, discovered a decade ago [\onlinecite{Hasan2010,Ando2013,Qi2011}], has led to flourishing of novel and fascinating physics. The simplicity of their surface states, the robustness of both topological properties and surface metallicity under external perturbations, and the prediction of novel quantized states arising from the peculiar coupling between magnetic and electric fields make TIs a perfect venue for the next generation electronic and spintronic devices.

Focusing on the spintronics aspect, one of the possibilities to obtain spin-polarized electronic currents is to use the spin splitting in momentum space of surface states that results from the breaking of inversion symmetry at a surface (Rashba effect) [\onlinecite{Rashba1960}]. This has recently stimulated studies of the electronic structure of~Bi and Bi-based compounds, because of the large spin-orbit interaction in this heavy group-V semimetal (see, e.g., Ref.~[\onlinecite{Hofmann2006}]). Along this line many investigations of the changes in the electronic structure of Bi thin films deposited on semiconductors of with respect to bulk~Bi have been performed, e.g., Si\hkl(111) [\onlinecite{Takayama2011,Takayama2012}], SiC\hkl(0001) [\onlinecite{Huang2014}],  and highly-oriented pyrolytic graphite [\onlinecite{Kowalczyk2011}]. Such studies on Bi thin films are also of importance to the field of topological insulators [\onlinecite{Hasan2010,Ando2013,GaoChunLei2013,Drozdov2014,Takayama2015}].

As far as thin layers are concerned, it has been theoretically predicted that the Bi\hkl(111) surface could behave as a topological insulator [\onlinecite{Fu2007,Liu2011, Huang2013}]. Key factors leading to this property are tensile strength due to the mismatch between the lattice of the Bi deposit and that of the substrate, as well as the formation of chemical bonds at the Bi/substrate interface. Thus the investigation of such interfaces is a great relevance to explore the existence of new physical properties and, if any, to open ways to engineer the relevant interfaces. On the experimental side, III-V substrates have not been explored up to now. We report here on a study of the electronic structure of Bi films deposited on both A (In-terminated) and B (As-terminated) faces of InAs\hkl(111). Although some results on the growth of Bi on InAs\hkl(111)B have been reported [\onlinecite{Szamota-Leandersson2011}], to our knowledge that of Bi on InAs\hkl(111)A has not yet been investigated.

We have followed the structural and electronic changes that occur during the deposition of Bi on both faces of InAs\hkl(111) using low-energy electron diffraction (LEED), 
scanning tunneling microscopy (STM), 
and core-level as well as valence-band photoelectron spectroscopy (PES) using synchrotron radiation. The growth mode is found to be dependent on the substrate side, an epitaxial growth being evidenced on the A~side. As a consequence, ultra-thin Bi films deposited on InAs\hkl(111)A have an the electronic structure showing all the characteristics of bulk~Bi. This does not occur when deposition is done on the B face because then the Bi growth proceeds via the formation of islands. The angle-resolved PES (ARPES) data on Bi/InAs\hkl(111)A are well explained by ab initio fully relativistic multiple scattering theory in the framework of the density functional theory.

\section{Experimental and computational methods}
\label{subsec:Experimental}

The experiments were performed at the APE beamline [\onlinecite{Panaccione2009}] of the Elettra synchrotron radiation facility (Trieste, Italy) and at the I3 beamline [\onlinecite{Balasubramanian2010}] of the MAX~III synchrotron (MAX-lab, Lund, Sweden). The substrates were cut from 0.5~mm thick $n$-doped ($3 \times 10^{18}$~cm$^{-3}$) InAs\hkl(111) wafers (Wafer Technology Ltd., UK) polished on both sides. After degassing, both InAs\hkl(111)A and B surfaces were prepared (simultaneously) by repeated cycles of ion bombardment (Ar$^{+}$, \SI{600}{\electronvolt}) and annealing at 400$^\circ$C until a sharp $(2 \times 2)$  or $(1 \times 1)$ LEED pattern for side~A or side~B, respectively, was obtained (see, e.g., Ref.~[\onlinecite{Richter2016}] and the references cited therein). Bismuth was deposited from a Knudsen cell at a rate of about \SI{0.5}{\BL\per\minute} under a pressure lower than~\SI{4e-10}{\Torr}.

On the theoretical side, band structure calculations were performed using the SPR-KKR package~[\onlinecite{Ebert2011}] that is based on the Korringa--Kohn--Rostoker method which uses the Green's function formalism within the multiple scattering theory. The package is based on the Dirac equation, thus fundamentally containing all relativistic effects, such as spin-orbit coupling. 
The calculations were performed within the local density and atomic sphere approximations for the bulk potential, that was used as an input to simulate the ARPES spectra within the one-step model of photoemission~[\onlinecite{Braun1996}]. The model of Rundgren--Malmstr\"{o}m [\onlinecite{Rundgren1977}] was used to connect the bulk potential with the surface. The relaxation of the surface was also taken into account using the structural data given in Ref.~[\onlinecite{Monig2005}].

\section{Results and discussion}
\label{sec:Results_Discussion}
\subsection{Pristine InAs\hkl(111) surfaces}
\label{subsec:Pristine_InAs}
\begin{figure}[h!]
\centering

\includegraphics[width=8.0cm]{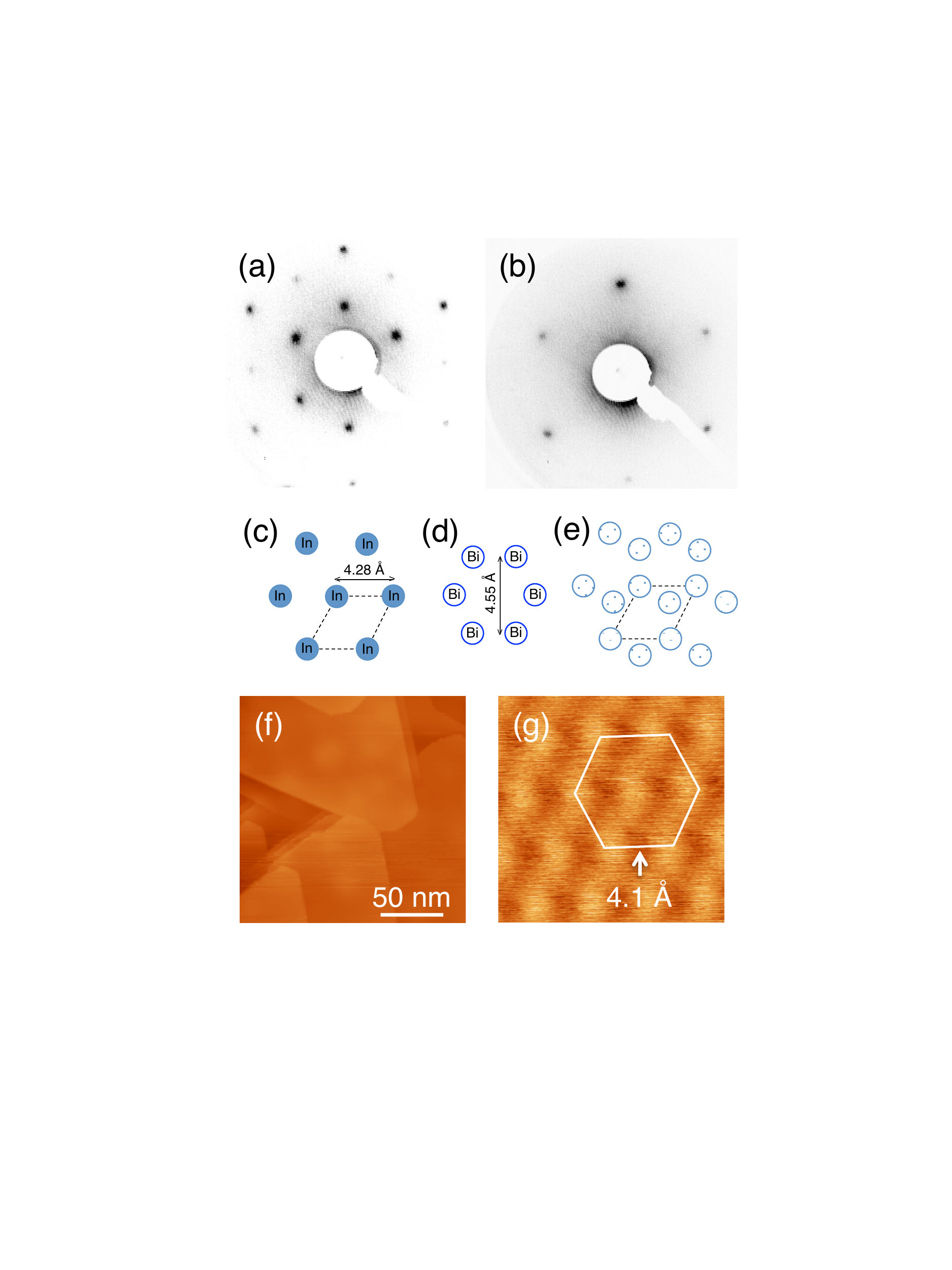}

\caption{ 
a) and b) LEED patterns of pristine InAs\hkl(111)A  showing the $(2 \times 2)$ reconstruction and of a \SI{10}{\BL}~Bi deposit InAs\hkl(111)A (taken at 34 and \SI{53}{\electronvolt}, respectively). Structures of (In-terminated) InAs\hkl(111)A c), Bi\hkl(111) d), and a possible arrangement of the Bi deposit on InAs e). (f and g) STM images of the InAs\hkl(111)A substrate covered by \SI{10}{\BL}~Bi.}
\label{fig:LeedALS_STM_VESTAsht.pdf}
\end{figure}

Although a $(2 \times 2)$ LEED pattern is observed for InAs\hkl(111)A [Fig.~\ref{fig:LeedALS_STM_VESTAsht.pdf} (a)], the InAs\hkl(111)B surface is unreconstructed (not shown); this is due to the presence of In vacancies on the In-terminated A surface (see Refs.~[\onlinecite{Taguchi2006, Taguchi2005, Hilner2010}], and the references cited therein). 

\begin{figure}
\centering
\includegraphics[width=7.00cm]{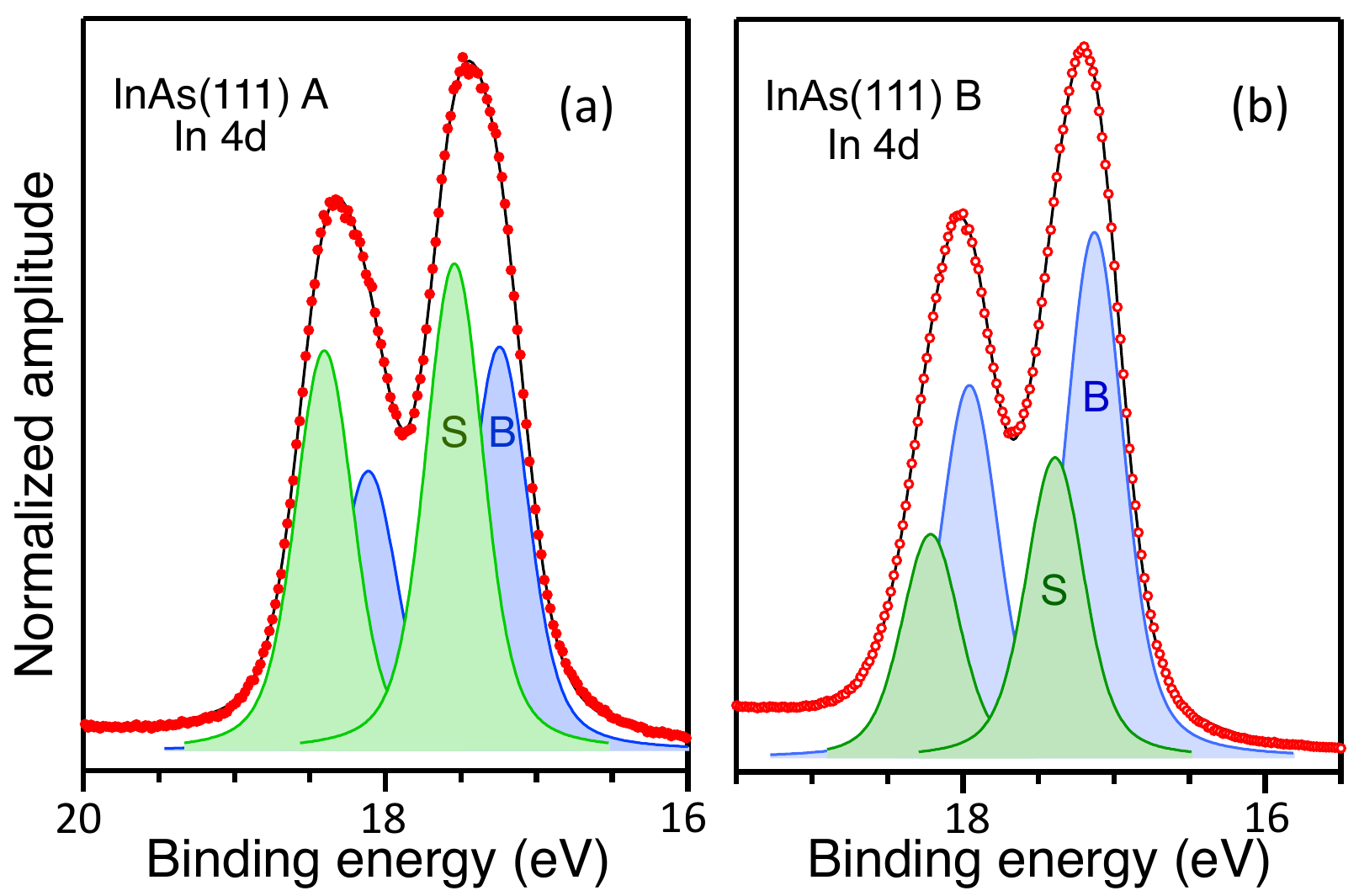}
\caption{In~$4d$ PES spectra of clean InAs\hkl(111): (a) A~side; (b) B~side. Raw data are shown as dots; the lines are the fit to the data and the components involved in the fit.}
\label{fig:Fig2.pdf}
\end{figure}

The In~$4d$ PES spectra for both the A side and the B side of InAs\hkl(111) are in agreement with those reported previously [\onlinecite{Andersson1994, Szamota-Leandersson2003, Szamota-Leandersson2011}]. They are given in Fig.~\ref{fig:Fig2.pdf} to help comparison in the following discussion with those of the Bi-covered substrates.

The origin of a $\approx \SI{250}{\milli\electronvolt}$ shift towards the high binding energy (BE) side of the main peak of the In~$4d$ PES spectrum for the A side with respect to the B side is due to an effective charge transfer perpendicular to the surface resulting in a polarization of the As-In bi-layers of InAs(111)B $(1 \times 1)$ [\onlinecite{Mankefors1999}].

It was recognize early ~[\onlinecite{Andersson1994}] that appart from the bulk component (denoted as~B) In $4d$ core levels contain as well a surface contribution (denoted as~S). Our experimental data are well reproduced fitting these two components by a Voigt profile with Gaussian and Lorentzian full width at half maximum of 0.18 eV
and 0.35 eV, respectively. One can also note that the intensity of the S component is larger for the In-terminated A side than for the B side. The B--S energy shift is found to be 0.3 eV for the A side, in good agreement with ~[\onlinecite{Andersson1994}].

\subsection{Bi/InAs\hkl(111): core levels}
\label{subsec:Bi_growth_AB}

At the early stage of the deposition, the LEED pattern of the InAs\hkl(111)A $(2 \times 2)$ substrate [Fig.~\ref{fig:LeedALS_STM_VESTAsht.pdf}(a)] starts to blur and relatively broad spots showing to the formation of Bi\hkl(111) layers appear. A well-defined sharp spots of Bi\hkl(111) $(1 \times 1)$ being observed for a Bi coverage greater than approximately \SI{10}{\BL}, see Fig.~\ref{fig:LeedALS_STM_VESTAsht.pdf}(b). 

This is consistent with the fact that the In-terminated InAs\hkl(111)A surface exhibits an hexagonal pattern of In atoms separated by $\approx \SI{4.28}{\angstrom}$ lattice [Fig.~\ref{fig:LeedALS_STM_VESTAsht.pdf}(c)] and the Bi\hkl(111) an honeycomb pattern in which the distance between a Bi atom and its second in-plane neighbor being $\approx \SI{4.55}{\angstrom}$ [Fig.~\ref{fig:LeedALS_STM_VESTAsht.pdf}(d)]. This $\approx 6\%$ mismatch is small enough to allow an epitaxial growth of Bi\hkl(111) on the InAs\hkl(111) substrate, as shown for instance in the manner depicted in Fig.~\ref{fig:LeedALS_STM_VESTAsht.pdf}(e) which is compatible with the LEED patterns in Figs.~\ref{fig:LeedALS_STM_VESTAsht.pdf}(a) and (b).

\begin{figure}[t!]
\centering

\includegraphics[width=7.0cm]{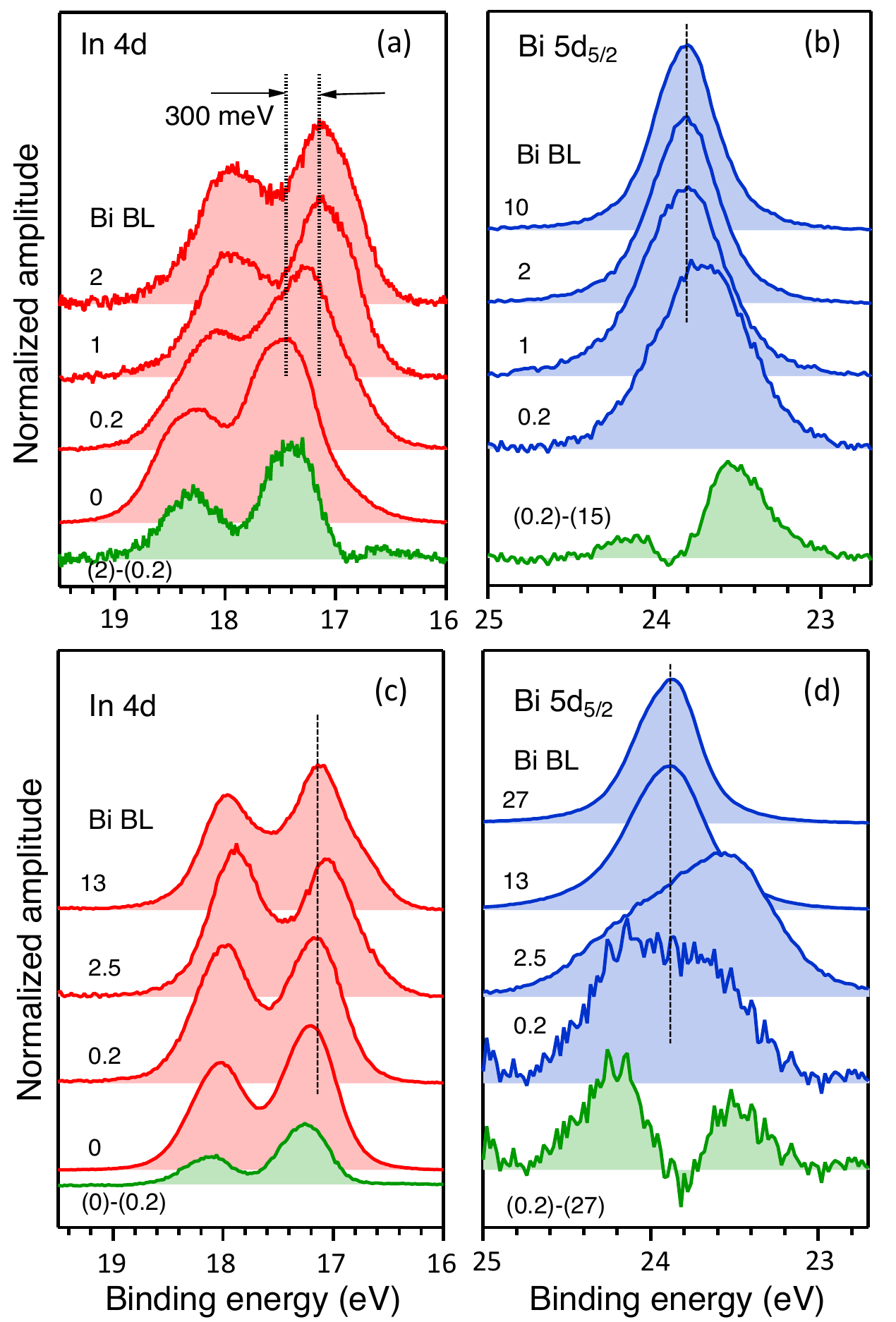}
\caption{Changes in core level PES spectra upon Bi deposition (in~BL) on InAs\hkl(111)A and B (top and bottom panels, respectively): (a)~In~$4d$ spectra. The difference spectrum between In~$4d$ PES signals for 2.0 and \SI{0.2}{\BL} Bi coverage is shown (in green) at the bottom of the figure. (b)~Bi~$5d_{5/2}$ spectra. The difference spectrum between Bi~$5d_{5/2}$ PES signals for 0.2 and \SI{15}{\BL} Bi coverage is shown (in green) at the bottom of the figure. (c)~In~$4d$; the difference spectrum between In~$4d$ PES signals for 0 and \SI{0.2}{\BL} Bi deposition is shown (in green) at the bottom of the figure; (d)~Bi~$5d_{5/2}$ spectra, the difference spectrum between Bi~$5d_{5/2}$ PES signals for 0.2 and \SI{27}{\BL} Bi deposition is shown (in green) at the bottom of the figure.}

\label{fig3_4merged.pdf}
\end{figure}

The evolution of the intensity of the Bi $5d$ and In $4d$ PES signals as a function of the Bi evaporation duration on InAs\hkl(111)A (not shown) is made of a series of linear segments with break points, separated by a common evaporation duration, where a change in slope occurs (see top panel of Fig.~\ref{fig3_4merged.pdf}). This is the well-known signature of a layer-by-layer (Frank--van der Merwe) growth mode. 
This is consistent with both the LEED patterns and the scanning tunneling microscopy (STM) images [Fig.~\ref{fig:LeedALS_STM_VESTAsht.pdf}(d)]
of these surfaces which terraces with the six-fold symmetry [Fig.~\ref{fig:LeedALS_STM_VESTAsht.pdf}(c)] characteristic of  the hexagonal arrangements of the Bi atoms on {\hkl(111) terraces}  [Fig.~\ref{fig:LeedALS_STM_VESTAsht.pdf}(d)] [\onlinecite{Richter2016}].

On the InAs\hkl(111)B surface the plot of the intensity of the Bi~$5d$ and In~$4d$ PES signals as a function of the Bi deposition duration shows a break in linear behavior only for the first \SI{1}{\BL} Bi. 
This suggests that the Bi growth on the B side of InAs\hkl(111) proceeds via a Stranski--Krastanow mode, contrary to what happens on InAs\hkl(111)A. This is confirmed by the fact that an amount of $\sim \SI{50}{\BL}$~Bi is necessary to clear the substrate In~$4d$ PES signal, i.e., to ensure the coalescence of the Bi three-dimensional islands that build on the surface after one Bi BL has been deposited on the B side of~InAs.

Changes in the electronic structure of the Bi/InAs\hkl(111)A interface can be detected at the early stage of the Bi deposition in the In~$4d$ PES spectrum (recorded at $h\nu = 45$~eV) [Fig.~\ref{fig3_4merged.pdf}(a)]. For \SI{0.2}{\BL} coverage the whole spectrum shifts relative to that of the pristine substrate by $\approx \SI{200}{\milli\electronvolt}$ towards the high BE side; this shift reaches $\approx \SI{300}{\milli\electronvolt}$ for a coverage of $\approx \SI{1}{\BL}$ and stabilizes to this value for higher coverages.
This shift is accompanied by a change in the overall shape of the In~$4d$ spin-orbit doublet. Both effects are due to the progressive disappearance of the surface S component present in the spectrum of pristine InAs\hkl(111)A upon Bi deposition, see Fig.~\ref{fig:Fig2.pdf}. The substrate In~$4d$ PES signal disappears totally for a Bi coverage of $\approx \SI{10}{\BL}$.

The bottom spectrum in Fig.~\ref{fig3_4merged.pdf}(a) presents the subtraction between \SI{2}{\BL} and   \SI{0.2}{\BL} Bi coverages. This difference spectrum is very similar to that corresponding to the S component in the In~$4d$ PES spectrum of pristine InAs\hkl(111)A. This is thus a confirmation of the disappearance of the S component at the early stage of Bi deposition on InAs\hkl(111)A. Note that we did not subtract the In~$4d$ PES for \SI{0.2}{\BL} Bi coverage because of the band bending effect induced by the deposition of the first ~Bi atoms.
Finally it is interesting to note that the subtraction reveals a faint structure at $\approx \SI{17.4}{\electronvolt}$ BE, which corresponds to the BE value expected when Bi is chemically bound to~In. Such a behavior is also observed for the Bi/GaAs\hkl(111)A and B interfaces [\onlinecite{McGinley1999,McGinley2000}] as well as for the Bi/InAs\hkl(100) interface [\onlinecite{Szamota-Leandersson2009,Ahola-Tuomi2011}]. 

We now turn to the Bi~$5d_{5/2}$ PES signal [Fig.~\ref{fig3_4merged.pdf}(b)]. When the amount of Bi deposited is increased its shape converges to that of bulk Bi at $\approx \SI{10}{\BL}$ of Bi.  
To shed light on the origin of the changes occurring at low Bi coverage we show at the bottom of Fig.~\ref{fig3_4merged.pdf}(b) the difference spectrum obtained by subtracting the Bi~$5d_{5/2}$ PES spectrum for \SI{15}{\BL} from that for \SI{0.2}{\BL} Bi coverage, removing in such a way the bulk component. This difference spectrum is made of two components: the one at low BE, of much higher intensity, can be attributed to Bi bound to In and that at high BE to Bi bound to As (see Ref.~[\onlinecite{Szamota-Leandersson2011}]). The latter can be present despite the InAs\hkl(111)A substrate is In-terminated because some defects due to surface preparation (ion bombardment and annealing) might allow As atoms to be present on the surface. The subtraction clearly indicates that the apparent change in the position of the Bi~$5d_{5/2}$ PES upon Bi deposition, namely for \SI{0.2}{\BL} Bi coverage, arises from the intensity variation/decrease of the PES component corresponding to the Bi--In bond, similarly to what is observed for the In~$4d$ PES signal upon Bi deposition.

The evolution of In~$4d$ and Bi $5d_{5/2}$ PES spectra on InAs\hkl(111)B is quite different from InAs\hkl(111)A [Fig.~\ref{fig3_4merged.pdf}(c) and (d)]. The presence of Bi atoms on the B surface does not induce a shift as for InAs\hkl(111)A. This allows a subtraction of the In $4d$ spectrum of the bare substrate from that corresponding to \SI{0.2}{\BL} Bi deposition ["(0)-(0.2)", at the bottom of Fig.~\ref{fig3_4merged.pdf}(c)]. Similarly to InAs\hkl(111)A, the S component disappears at very early stages of Bi deposition. For thicker Bi layers a new low BE feature appears in the spectra. 

A better insight of its evolution is shown in Fig.~\ref{fig3_4merged.pdf}(a) where we plot only difference spectra. The surface component, "(0)-(0.2)" from Fig.~\ref{fig3_4merged.pdf}(c), is sh for own for a reference. Subtracting In $4d$ spectrum corresponding to \SI{0.2}{\BL} Bi and containing only the bulk component, from that of \SI{13}{\BL} and \SI{27}{\BL} Bi depositions, middle and upper spectrum, respectively (only In $4d_{5/2}$ component is shown), reveals two new structures that can be attributed to Bi-In (higher BE) and In-In (lower BE) components.
Such a feature has been previously suggested [\onlinecite{Szamota-Leandersson2011}] but not resolved. Here, the In-In component clearly appears for higher Bi coverages. We believe that its formation is due to the fact that the Bi-As bond is stronger as compared to the Bi-In one, as deduced from BEs shifts  [\onlinecite{Szamota-Leandersson2011}]. At higher coverages Bi-As bonds being formed, releasing free In atoms on the surface. Similar chemical behavior has been observed when annealing Bi layer on top of InAs(111) crystal [\onlinecite{Richter2016}]. 

\begin{figure}[t!]
\centering
\includegraphics[width=8.0cm]{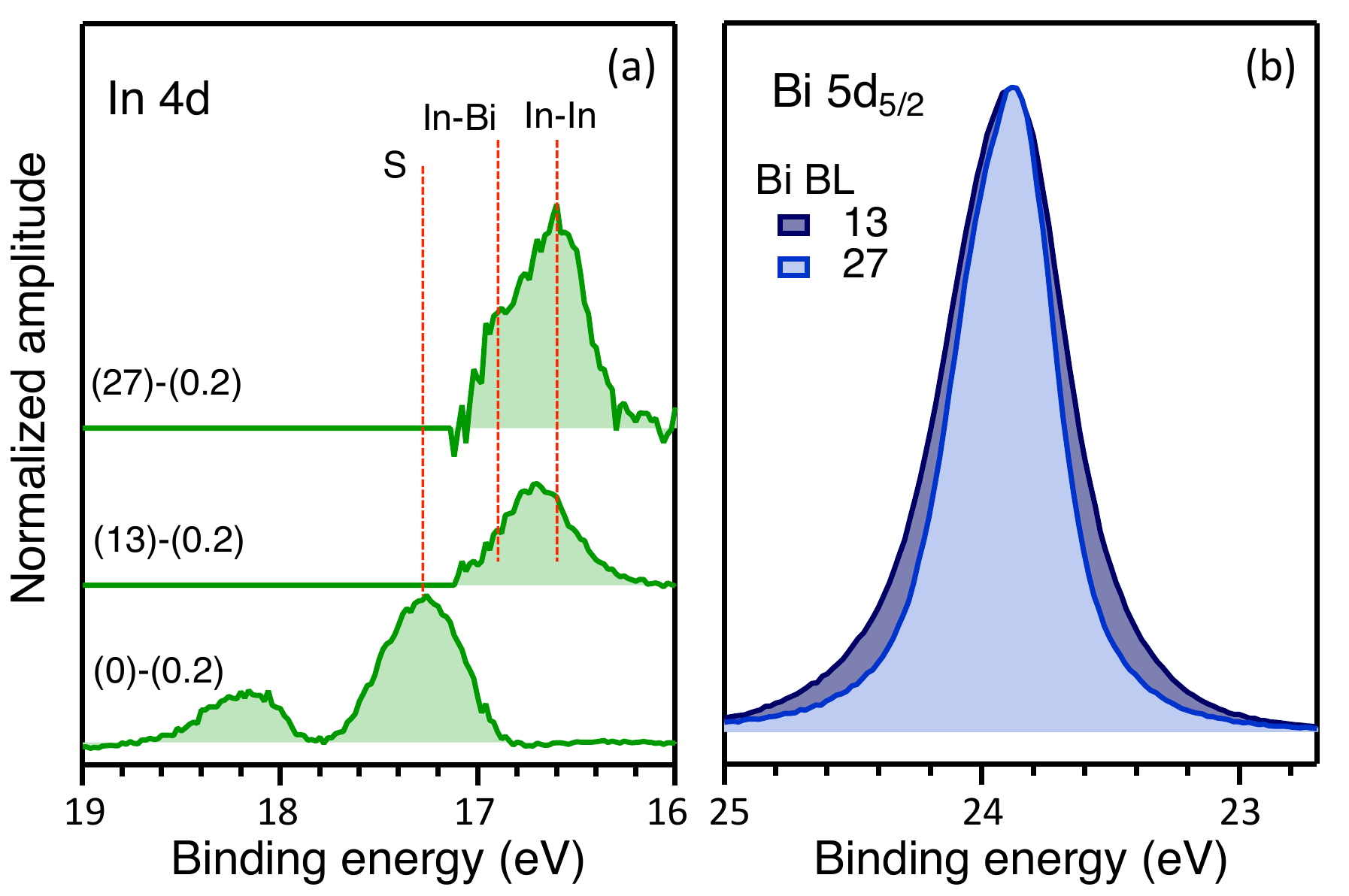}
\caption{(Color online) PES difference spectra between Bi-covered (in BL) and pristine InAs\hkl(111)B: (a) In~$4d$; (b) Bi~$5d_{5/2}$.}
\label{fig:Fig5.pdf}
\end{figure}

Evolution of Bi $5d_{5/2}$ PES spectra [Fig.~\ref{fig3_4merged.pdf}(d)] are in agreement with this chemical interpretation. As in the case of InAs(111)A side we show the difference spectrum [bottom in Fig.~\ref{fig3_4merged.pdf}(d)] obtained by subtracting the Bi $5d_{5/2}$ PES spectrum for \SI{0.2}{\BL} Bi coverage from that of \SI{13}{\BL} coverage. In the difference spectrum we find again two components, the one at high BE has now a larger intensity which is expected for the As-terminated B~side. The presence of the low BE component, attributed to Bi-In, can be justified because the annealing during the sample preparation eliminates preferentially As atoms leaving In-terminated patches on the surface. This image is consistent with further evolution of Bi $5d_{5/2}$ spectra. For \SI{2.5}{\BL} Bi coverage, as all Bi-As bonds are saturated, the Bi-In is a dominant component. Upon increasing deposition the FHWM of Bi $5d_{5/2}$ decreases and reaches that of bulk Bi well above 27~BL [Fig.~\ref{fig3_4merged.pdf}(b)], when Bi islands coalesce in larger terraces, in clear contrast with InAs(111)A surface. We suggest that the presence of strong Bi-As bonds prevents epitaxial growth at very initial stages of Bi deposition.

Reminding that for such high depositions the In~$4d$ PES signal is faint but still present, this is in line with the formation of 3D Bi islands during the growth. This situation is similar to that encountered on GaAs where \emph{Bi bands} have been reported for a \SI{2}{\BL} deposit [\onlinecite{McGinley1999}]. Indeed the STM images show the presence of such islands on InAs\hkl(111)B [\onlinecite{Hilner2010}].

We deposited Bi films as well on MBE prepared surfaces. Interestingly, in spite of their higher quality, as compared to the ion-bombardment/annealing preparation, Bi growth on the InAs(111)A face is worse. In MBE systems, it is impossible to suppress instantaneously the As background pressure  after the InAs substrate growth. So, even the perfect In-terminated A-face is covered by a fraction of monolayer of As preventing the good epitaxy. Paradoxically, Bi monocrystals prepared on bombarded/annealed A-face are of much higher quality.

\begin{figure}
\centering
\includegraphics[width=8.5 cm]{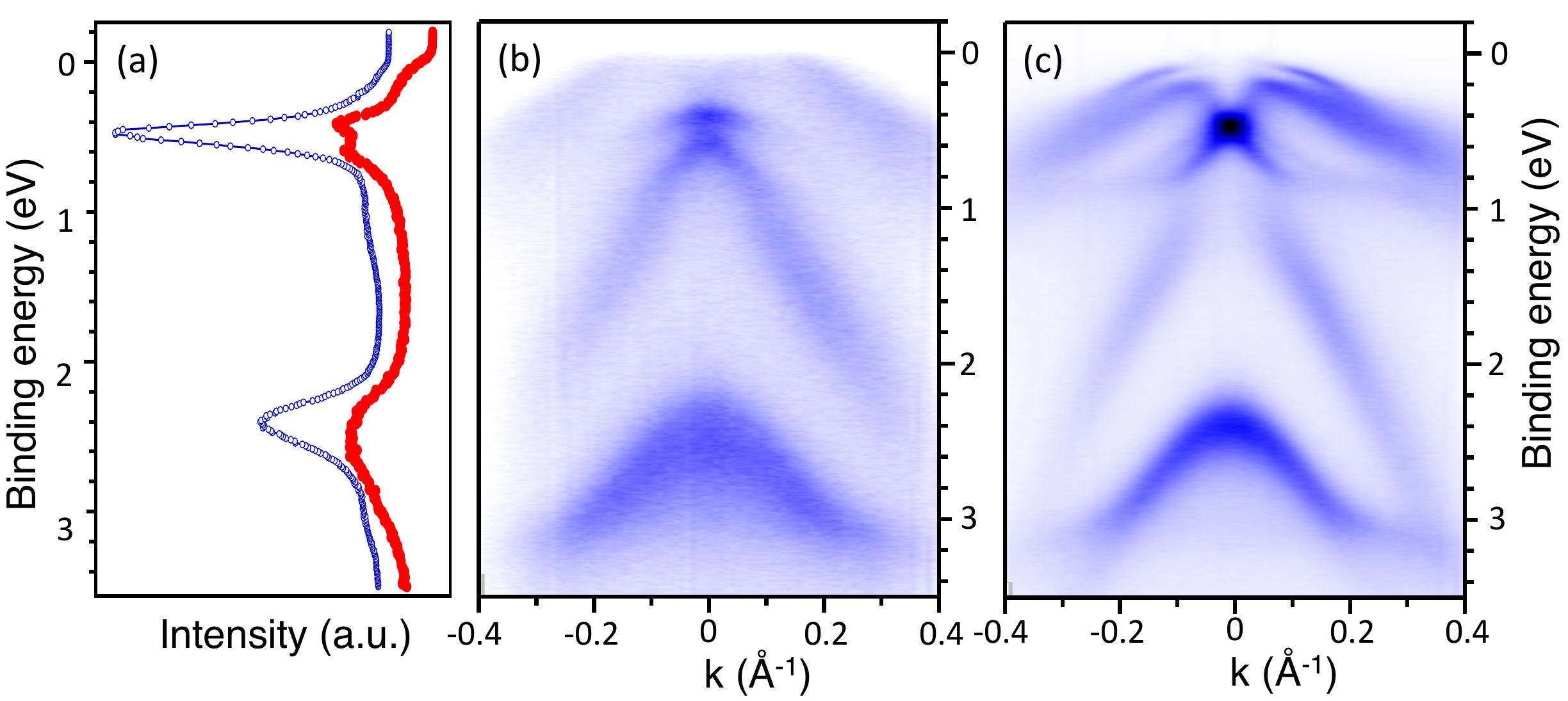}
\caption{(Color online) ARPES spectra along the $\overline{\Gamma}$--$\overline{\mathrm{K}}$ line of InAs\hkl(111)A covered by (b) \SI{2}{\BL} Bi and (c) \SI{12}{\BL} Bi. (a)~Normal emission EDC corresponding to (b) (red) and~(c) (blue). Spectra were recorded at $h\nu = \SI{20}{\electronvolt}$}
\label{fig:Fig6.pdf}
\end{figure}

\subsection{Bi/InAs\hkl(111) valence band PES}
\label{subsec:Valence_band_PES_AB}

The ARPES spectra recorded at $h\nu = 20 $~eV along the $\overline{\Gamma}$--$\overline{\mathrm{K}}$ line for Bi  coverages of  \SI{2}{\BL} and \SI{12}{\BL}  are shown in Fig.~\ref{fig:Fig6.pdf}. A Bi coverage of \SI{2}{\BL}  is sufficient for the ARPES spectrum to show main bands characteristic of bulk Bi [Fig.~\ref{fig:Fig6.pdf}(b)], in particular Bi states near the Fermi level  ($E_{\mathrm{F}}$) start to appear in the gap of InAs. We observed that the bands sharpen progressively upon increasing Bi coverage, the whole ARPES spectrum  becoming identical to that of bulk Bi [\onlinecite{Hofmann2006}]  for \SI{12}{\BL} coverage [Fig.~\ref{fig:Fig6.pdf}(c)]. Normal emission spectra in Fig.~\ref{fig:Fig6.pdf}(a), extracted from Fig.~\ref{fig:Fig6.pdf}(b) and (c), show namely strong intensity increase of the surface state resonance at binding energy of 2 eV that characterize high quality Bi crystal [\onlinecite{Hofmann2006}]. 
This is in complete consistency with the Bi epitaxial growth mode on the A side, as deduced from the analysis of the core level PES spectra (see Sec.~\ref{subsec:Bi_growth_AB}).

\begin{figure}
\includegraphics[width=8.0 cm]{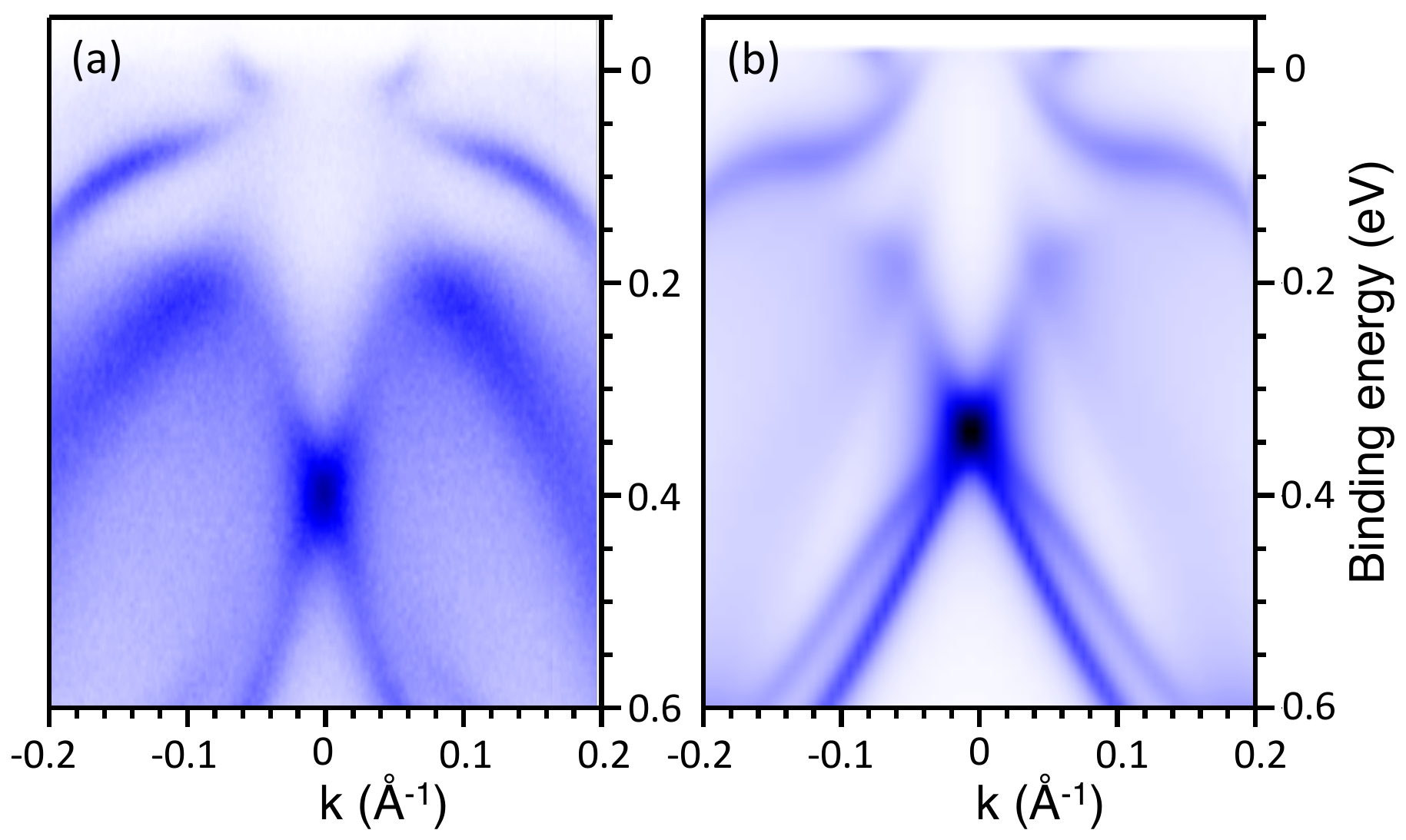}

\caption{(Color online) ARPES spectrum along the $\overline{\Gamma}$--$\overline{\mathrm{K}}$  direction of a Bi(111) monocrystal grown on an InAs(111)-A substrate (a) measured and (b) computed with the SPR-KKR package. The photon energies used were $h\nu = \SI{20}{\electronvolt}$ for the measurement and $h\nu = \SI{22}{\electronvolt}$ for the calculation with, for both cases, a circularly polarized light.}

\label{fig:Fig7.pdf}
\end{figure}

The measured spectrum was as well compared to calculations [see Fig.~\ref{fig:Fig7.pdf}]. Bi band dispersion calculations were carried out with the SPR-KKR package. The following lattice parameters were used: a = 4,5332 \AA ~and c = 11,7967 \AA.
The measured spectrum [Fig.~\ref{fig:Fig7.pdf}(a)] in the vicinity of the $\overline{\Gamma}$ point along the $\overline{\Gamma}$--$\overline{\mathrm{K}}$ direction are well reproduced by
the corresponding theoretical calculation [Fig.~\ref{fig:Fig7.pdf}(b)]. The theoretical simulation was performed for a Bi semi-infinite crystal with the [111] direction. The following ARPES calculations are based on the one-step model of photo-emission applied to a bulk potential; this approximation appears to be sufficient to reproduce the measured spectra. As seen from Fig.~\ref{fig:Fig7.pdf} the calculated bands are in good agreement with the calculation.

\begin{figure}
\includegraphics[width=8.5 cm]{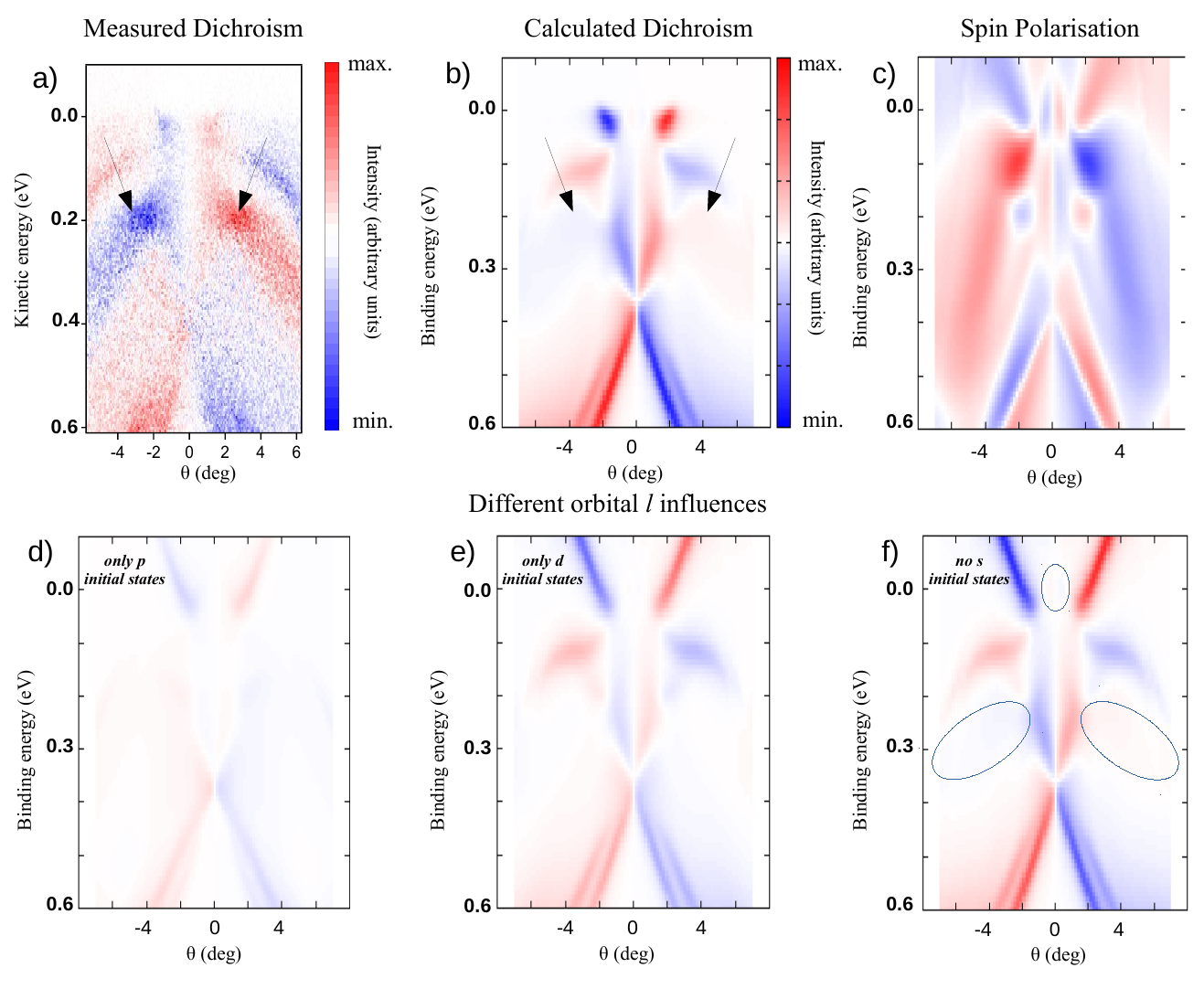}

\caption{(Color online) ARPES dichroic spectra of Bi\hkl(111): (a) Experiment ($h\nu =\SI{20}{\electronvolt}$) for \SI{20}{\BL} of Bi deposited on InAs\hkl(111)A. (b) Theory ($h\nu =\SI{22}{\electronvolt}$). (c) Calculated spin polarization. (d, e, and f) Theoretical analysis of the contribution of the angular momentum in  the initial state (respectively, $\ell = 1$, 2, and in absence of $\ell = 0$) to the dichroic response.}
\label{fig:Dichro1sht.pdf}
\end{figure}

The electronic bands show a high sensitivity to the helicity of the light used to probe them as shown by the PES dichroic spectrum recorded at $h\nu =\SI{20}{\electronvolt}$ in the $\overline{\Gamma}$--$\overline{\mathrm{M}}$ direction [Fig.~\ref{fig:Dichro1sht.pdf}(a)]. The theoretical prediction at $h\nu =\SI{22}{\electronvolt}$ [Fig.~\ref{fig:Dichro1sht.pdf}(b)] reproduces nicely the experimental result. These results can be compared to the spin polarization to display the correlation between the observed circular dichroism and the fundamental spin polarization. 

In addition the theoretical analysis of the contribution to the dichroic signal of initial state having selected angular momentum is shown in Figs.~\ref{fig:Dichro1sht.pdf}(d--f). When only $p$ [Fig.~\ref{fig:Dichro1sht.pdf}(d)] or $d$ [Fig.~\ref{fig:Dichro1sht.pdf}(e)] initial states are considered  not all the bands experimentally observed are recovered. The remaining states, circled in blue, are regained when both $p$ and $d$ symmetries are considered as seen in Fig.~\ref{fig:Dichro1sht.pdf}(d)] where only the $s$ orbitals were deactivated. This suggests a partial hybridization of the $p$ and $d$ bands.

At this stage it deserves to be reminded that circular dichroism in ARPES has been recently proposed as an alternative to spin-resolved ARPES because it is a much less experimentally experimental technique (see, e.g., Refs.~[\onlinecite{Wang2011,Park2012,Wang2013}]). However, extensive combined experimental and theoretical studies of ARPES using circularly polarized light have demonstrated that contrasted experimental findings between spin-resolved and dichroic ARPES arise from the spin dependence of the relativistic dipole matrix elements that also depend strongly on the final states reached in the PES, i.e., on the photon energy inducing the photoelectron. This underlines the fact that the final state involved the photoemission process play a role that must be more influential on the result than the spin texture of the initial state. Actually, sign inversions are observed rather regularly when increasing the photon energy by several~eV and coincide only accidentally with the spin polarization [see Fig.~\ref{fig:Dichro1sht.pdf}(c)].

\section{Conclusions}

We show that InAs(111) substrates prepared by two to three cycles of successive ion sputtering and annealing present distinctive results for the Bi growth depending on the atomic surface termination. Core-level analysis confirm that Bi growth is epitaxial on the InAs(111)-A surface even for very thin films. Deposition of 10 bilayers of Bi on InAs(111)-A resulted in a Bi monocrystal of very high quality as attested by reproduction of the band dispersion by theoretical calculations. Moreover, a close comparison of circular dichroism spectra between theory and experiment highly suggests a pd-orbitals hybridization. 

\begin{acknowledgments}
The research leading to these results has received funding from the European Community's Seventh Framework Programme (FP7/2007-2013) under grant agreement~n$^{\circ}$312284. 
\end{acknowledgments}


%

\end{document}